\documentclass[a4paper]{spie}  

\newcommand{\Hitomi}{\textit{Hitomi}} 
 
\newcommand{\Msun}{$M_{\odot}$}
 
\usepackage{amsmath,amsfonts,amssymb}
\usepackage[dvipdfmx]{graphicx}
\usepackage[colorlinks=true, allcolors=blue]{hyperref}
\usepackage{aas_macros}
\usepackage{wrapfig}

\title{A broadband X-ray imaging spectroscopy in the 2030s: the FORCE mission}

\author[a]{Koji Mori}
\author[b]{Takeshi G. Tsuru}
\author[c]{Kazuhiro Nakazawa}
\author[d]{Yoshihiro Ueda}
\author[e]{Shin Watanabe}
\author[f]{Takaaki Tanaka}
\author[e]{Manabu Ishida}
\author[g]{Hironori Matsumoto}
\author[h]{Hisamitsu Awaki}
\author[i]{Hiroshi Murakami}
\author[j]{Masayoshi Nobukawa}
\author[a]{Ayaki Takeda}
\author[k]{Yasushi Fukazawa}
\author[g]{Hiroshi Tsunemi}
\author[l]{Tadayuki Takahashi}
\author[m]{Ann Hornschemeier}
\author[m]{Takashi Okajima}
\author[m]{William W. Zhang}
\author[m]{Brian J. Williams}
\author[m]{Tonia Venters}
\author[m,n]{Kristin Madsen}
\author[m,o]{Mihoko Yukita}
\author[p]{Hiroki Akamatsu}
\author[q]{Aya Bamba}
\author[r]{Teruaki Enoto}
\author[s]{Yutaka Fujita}
\author[t]{Akihiro Furuzawa}
\author[u]{Kouichi Hagino}
\author[v]{Kosei Ishimura}
\author[w]{Masayuki Itoh}
\author[x]{Tetsu Kitayama}
\author[y]{Shogo Kobayashi}
\author[z]{Takayoshi Kohmura}
\author[aa]{Aya Kubota}
\author[d]{Misaki Mizumoto}
\author[k]{Tsunefumi Mizuno}
\author[u]{Hiroshi Nakajima}
\author[ab]{Kumiko K. Nobukawa}
\author[g]{Hirofumi Noda}
\author[q]{Hirokazu Odaka}
\author[ac]{Naomi Ota}
\author[ad]{Toshiki Sato}
\author[h]{Megumi Shidatsu}
\author[f]{Hiromasa Suzuki}
\author[k]{Hiromitsu Takahashi}
\author[af]{Atsushi Tanimoto}
\author[ae]{Yukikatsu Terada}
\author[h]{Yuichi Terashima}
\author[b]{Hiroyuki Uchida}
\author[ad]{Yasunobu Uchiyama}
\author[e]{Hiroya Yamaguchi}
\author[ag]{Yoichi Yatsu}

\affil[a]{Department of Applied Physics and Electronic Engineering, University of Miyazaki, Miyazaki 889-2192, Japan}
\affil[b]{Department of Physics, Kyoto University, Kyoto 606-8502, Japan}
\affil[c]{Department of Physics, University of Tokyo, Tokyo 113-0033, Japan}
\affil[d]{Department of Astronomy, Kyoto University, Kyoto 606-8502, Japan}
\affil[e]{Institute of Space and Astronautical Science (ISAS), Japan Aerospace Exploration Agency (JAXA), Kanagawa 252-5210, Japan}
\affil[f]{Department of Physics, Konan University, Hyogo 658-8501, Japan}
\affil[g]{Department of Earth and Space Science, Osaka University, Osaka 560-0043, Japan}
\affil[h]{Department of Physics, Ehime University, Ehime 790-8577, Japan}
\affil[i]{Department of Information Science, Faculty of Liberal Arts, Tohoku Gakuin University, Miyagi 981-3193, Japan}
\affil[j]{Department of Teacher Training and School Education, Nara University of Education, Nara 630-8528, Japan}
\affil[k]{Department of Physical Science, Hiroshima University, Hiroshima 739-8526, Japan}
\affil[l]{Kavli Institute for the Physics and Mathematics of the Universe (WPI), The University of Tokyo, Chiba, 277-8583, Japan}
\affil[m]{NASA's Goddard Space Flight Center, Greenbelt, MD 20771, USA}
\affil[n]{CRESST, Department of Physics, and Center for Space Science and Technology, UMBC, Baltimore, MD 21250, USA}
\affil[o]{The Johns Hopkins University, Homewood Campus, Baltimore, MD 21218, USA}
\affil[p]{SRON Netherlands Institute for Space Research, 3584 CA Utrecht, The Netherlands}
\affil[q]{Department of Physics, The University of Tokyo, Tokyo, 113-0033, Japan}
\affil[r]{Extreme natural phenomena RIKEN Hakubi Research Team, Cluster for Pioneering Research, RIKEN, Saitama 351-0198, Japan}
\affil[s]{Department of Physics, Tokyo Metropolitan University, Tokyo 192-0397, Japan}
\affil[t]{Fujita Health University, Aichi 470-1192, Japan}
\affil[u]{College of Science and Engineering, Kanto Gakuin University, Kanagawa 236-8501, Japan}
\affil[v]{Faculty of Science and Engineering, Waseda University, Tokyo 169-8555, Japan}
\affil[w]{Department of Human Environmental Science, Kobe University, Hyogo 657-8501, Japan}
\affil[x]{Department of Physics, Toho University, Chiba 274-8510, Japan}
\affil[y]{Department of Physics, Tokyo University of Science, Tokyo 162-8601, Japan}
\affil[z]{Department of Physics, Tokyo University of Science, Chiba 278-8510, Japan}
\affil[aa]{Department of Electronic Information Systems, Shibaura Institute of Technology, Saitama 337-8570, Japan}
\affil[ab]{Department of Science, Kindai University, Osaka 577-8502, Japan}
\affil[ac]{Department of Physics, Nara Women's University, Nara 630-8506, Japan}
\affil[ad]{Department of Physics, Rikkyo University, Tokyo 171-8501, Japan}
\affil[ae]{Graduate School of Science and Engineering, Saitama University, Saitama, 338-8570, Japan}
\affil[af]{Graduate School of Science and Engineering, Kagoshima University, Kagoshima, 890-8580, Japan}
\affil[ag]{Department of Physics, Tokyo Institute of Technology, Tokyo 152-8551, Japan}

\authorinfo{Further author information: (Send correspondence to K.M.)\\K.M.: E-mail: mori@astro.miyazaki-u.ac.jp, Telephone: 81 985 58 7371}
 
\begin{document} 
\maketitle

\begin{abstract}
In this multi-messenger astronomy era, all the observational probes are improving
their sensitivities and overall performance. The Focusing on Relativistic universe
and Cosmic Evolution (FORCE) mission, the product of a JAXA/NASA collaboration, will
reach a 10 times higher sensitivity in the hard X-ray band ($E >$ 10~keV) in
comparison with any previous hard X-ray missions, and provide simultaneous soft
X-ray coverage. FORCE aims to be launched in the early 2030s, providing a perfect
hard X-ray complement to the ESA flagship mission Athena. FORCE will be the most
powerful X-ray probe for discovering obscured/hidden black holes and studying high
energy particle acceleration in our Universe and will address how relativistic
processes in the universe are realized and how these affect cosmic evolution. FORCE,
which will operate over 1--79 keV, is equipped with two identical pairs of
supermirrors and wideband X-ray imagers. The mirror and imager are connected by a
high mechanical stiffness extensible optical bench with alignment monitor systems
with a focal length of 12~m. A light-weight silicon mirror with multi-layer coating
realizes a high angular resolution of $<15''$ in half-power diameter in the broad
bandpass. The imager is a hybrid of a brand-new SOI-CMOS silicon-pixel detector and
a CdTe detector responsible for the softer and harder energy bands,
respectively. FORCE will play an essential role in the multi-messenger astronomy in
the 2030s with its broadband X-ray sensitivity.
\end{abstract}


\section{INTRODUCTION}
\label{sec:intro}  

What is the ultimate goal of astrophysics? The simplest answer is to understand how
the Universe works — how it started, how it has evolved into the present shape, how
it will ultimately be. We seek to learn what physical laws govern both the Universe
and its contents. These are big topics, requiring that we observe all aspects of the
Universe by all the methods we have available. The concept of our proposing mission
is to focus on the high energy processes that govern much of the structure formation
and evolution of the Universe by means of highly sensitive wideband X-ray
observations.

High energy phenomena are ubiquitously seen at all levels of the hierarchical
structure of the Universe, from the largest gravitationally bound structures, galaxy
clusters, down to the scale of individual stars, in stellar explosions and in the
remnants of those explosions, including black holes. These high energy phenomena
play crucial roles in the structure formation and evolution of the Universe as we
can trace particle acceleration that affects how large structures form and the
supermassive black holes that are apparently coevolving with galaxies over cosmic
time.

These phenomena provide us with rich information concerning physical processes
working in the most extreme conditions, bringing dramatic deviations from
equilibrium states of the Universe. By understanding such extreme conditions, we
also verify the underlying physical conditions. Highly sensitive wideband X-ray
observations have a great advantage to study these high energy phenomena providing
sensitivity to both thermal emission from plasmas with temperatures of a few million
Kelvin (such as may be present in accretion disks around black holes) as well as
non-thermal emission (created by particle acceleration). There is a particular
opportunity at energies above 10 keV, where only very limited exploration has
occurred to date: highly-sensitive observations at these energies have the potential
to unveil the hidden Universe underneath thick, dense surrounding materials and/or
in the presence of overwhelming thermal emission that confuses observations at
energies below 10 keV.

The Focusing on Relativistic universe and Cosmic Evolution (FORCE)
mission\cite{2016SPIE.9905E..1OM, 2018SPIE10699E..2DN} is the product of a JAXA/NASA
collaboration and aims to be launched in the early 2030s.  FORCE will reach a 10
times higher sensitivity in the hard X-ray band ($E >$ 10 keV) in comparison with
any previous hard X-ray missions and provide simultaneous soft X-ray coverage. The
science goal of the FORCE mission is to understand how relativistic processes in the
universe are realized and how these affect cosmic evolution. In this paper, we
describe FORCE's scientific objectives, current design overview, and roles in
multi-messenger astronomy era.

\section{Scientific objectives}
\label{sec:objectivesAndRequirements}

We summarize the scientific goal into three scientific objectives as follows:
\begin{enumerate}
 \item How black holes grow and how they affect galaxy evolution,
 \item How non-thermal energy is generated and how much non- thermal energy is
       contained in the universe, and
 \item How stars evolve and explode.
\end{enumerate}

In this section, the three scientific objectives are reviewed.


\subsection{How black holes grow and how they affect galaxy evolution}

\subsubsection{Supermassive Black Hole: Co-evolution with their host galaxies}
\label{sec:AGN}

Nearly all galaxies in the universe contain supermassive black holes (SMBHs) in
their centers, whose masses range from $\sim$10$^{5}$ to 10$^{10}$ \Msun. Moreover,
a tight correlation between the masses of the central SMBHs and the masses of
galactic stellar bulges has been reported\cite{2009ApJ...698..198G}. These facts
indicate that the formation process of SMBHs and that of galaxies are strongly
coupled to each other, despite the large difference in their physical sizes ($>$8
orders of magnitude). Thus, it is one of the most fundamental questions in modern
astronomy how these SMBHs in galactic centers formed and ``co-evolved'' with their
host galaxies over the history of the Universe.

A key population for understanding the origin of the coevolution of galaxies and
SMBHs is the most heavily obscured Active Galactic Nuclei (AGNs), whose
line-of-sight hydrogen column density, $N_{\textrm H}$, is larger than $10^{24}$, so-called
Compton-thick AGNs (CTAGNs). According to the current best understanding, major
mergers of galaxies trigger violent star formation and intense mass accretion onto
SMBHs deeply ``buried'' by surrounding gas and dust. This means that a key aspect of
growth of SMBHs, the merging process, is largely hidden at most electromagnetic
wavelengths. Luckily, it is visible at energies above 10 keV. In fact, NuSTAR has
shown $>$50\% of late-stage AGN mergers in the local Universe are
Compton-thick\cite{2017ApJS..233...17R}, indicating that CTAGNs may be a distinct
population from normal AGNs, harboring much missing information about merger
growth. Thus, these CTAGNs (or buried AGNs) are unique tracers of the key processes
for understanding the coevolution, i.e., merger and subsequent rapid growth of
SMBHs. However, the basic questions on CTAGNs remain open: (1) how many CTAGNs,
largely missing from the current census, are present in the Universe and (2) what is
the contribution of CTAGNs to the total growth of SMBHs (merger, accretion, or
both?).

Hard X-ray surveys at energies above 10 keV provide the least biased AGN sample as
they have the strong penetrating power to overcome obscuration. However, due to
technical challenges, most of the previous X-ray surveys with good image quality
were performed in energy bands below 10 keV, which were insufficient to catch the
primary emission component from CTAGNs. There have been all-sky hard X-ray surveys
above $\sim$10 keV with non-focusing optics with Swift/BAT and INTEGRAL. These
surveys have found a part of the CTAGN population in the local universe. However,
their cosmological evolution requires a census to fainter populations with a true
focused imaging hard X-ray survey.

The first such imaging survey was conducted by the NuSTAR mission beginning in
2012. NuSTAR has arcminute angular resolution and has been an excellent pathfinder
for the FORCE mission, having resolved $\sim$35\% of the cosmic X-ray background
(CXB) in the 8--24 keV band\cite{2016ApJ...831..185H}. This CXB, X-ray emission that
is pervasive over the entire sky, represents the integrated emission from accreting
SMBHs over cosmic time. Buried within this signal, we expect to find a population of
CTAGNs. NuSTAR has provided us with new insights on CTAGN populations, implying that
there may be far more CTAGNs than those assumed in canonical AGN population
synthesis models\cite{2015ApJ...808..185C, 2017ApJ...846...20L}. However, much
deeper hard X-ray surveys, with $>$10 times better sensitivities than NuSTAR ($\sim
3 \times 10^{-15}~\textrm{erg}~\textrm{cm}^{-2}~\textrm{s}^{-1}$ in the 10--40 keV
band), must be conducted to determine the evolution of CTAGNs over cosmic time and
thus finally provide a complete picture of how SMBH growth occurs over the history
of the Universe.

\subsubsection{Intermediate-mass Black Hole: Nature of ultraluminous X-ray sources}

Central SMBHs in galaxies must have grown, via mergers and/or accretion, from
lighter ``seed'' black holes. Thus, another key population for understanding
formation mechanisms of SMBHs is that of intermediate-mass black holes (IMBHs),
those with masses of $\sim 10^{2\textrm{--}4}$~\Msun. The origin of IMBHs is
fundamentally important, connecting SMBHs to lower mass black holes: IMBHs may be
remnants of population III stars or even primordial black holes that formed just
after the Big Bang. We expect that some fraction of this IMBH population to remain
inside galaxies in the relatively local Universe today (e.g., those which did not
undergo growth via either merger or accretion to become SMBH). The gravitational
wave interferometers LIGO and VIRGO revealed the presence of black holes with masses
(up to 80~\Msun) considerably larger than ordinary stellar-mass ($\sim$5--15~\Msun)
black holes. However, to date, there is no truly convincing example of IMBHs in the
$10^{2\textrm{--}4}$~\Msun\ range. Identifying more IMBHs over a wide mass range
will provide strong constraints on theories of black hole evolution in the Universe.

When BHs accrete matter from their surroundings, they can be observed as luminous
X-ray emitters because the Eddington limit (the luminosity where the gravitational
force is balanced by radiation pressure) is nominally proportional to the black hole
mass. Ultra-Luminous X-ray sources (ULXs), off-nucleus compact X-ray sources found
in other galaxies with typical luminosities of
$10^{39\textrm{--}40}~\textrm{erg}~\textrm{s}^{-1}$ provide a useful population for
studying possible IMBHs (e.g., Ref.\citenum{2000ApJ...535..632M}). Although some
ULXs have turned out to harbor neutron stars\cite{2014Natur.514..202B,
2016ApJ...831L..14F} and many likely host Stellar-mass Black Holes (StMBHs), the
nature of the majority ($>$~400 sources including candidates;
Ref.\citenum{2011MNRAS.416.1844W}) is yet to be investigated with sufficiently
sensitive hard ($E >$~10 keV) X-ray observations because those are still too dim for
instruments currently available above 10~keV. While most ULXs may be stellar-mass
objects accreting at extreme rates which are also worth studying in terms of
accretion growth of black holes, it is possible that a handful of them are IMBHs
hidden within the ULX population. Of particular importance is to look for the hard
X- ray emission above 10 keV, as the peak hard X-ray luminosity has long been
proposed to be a BH-neutron star discriminator\cite{1996ApJ...473..963B}. We will
specifically search for hard X-ray emission from ULX sources in globular clusters
and satellite galaxies around giant elliptical galaxies as well as in dwarf
elliptical galaxies where the central black holes are likely to be of intermediate
mass. 


\subsubsection{Stellar-mass Black Hole: Missing population in our Galaxy}

To reveal the connection between the IMBHs and ``ordinary'' StMBHs produced by core
collapses of massive stars, we need to first establish statistical properties of
StMBHs, which are much less luminous than ULXs. The Milky Way galaxy is an excellent
target for these studies. It has been posited that 100 million StMBHs may have
formed within the Milky Way over its 10-billion-year
lifetime\cite{1998ApJ...496..155S}. Among them, only a few tens of bright
($L_{\mathrm X} > 10^{36}~\mathrm{erg~s}^{-1}$) X-ray binary systems that contain
black holes are currently known, most of which are
transients\cite{2016PASJ...68S...1N}. Thus, most of StMBHs in the Milky Way galaxy
are missing. The present sample is too small to constrain the true underlying mass
function and spatial distribution.

It is expected that StMBHs (either isolated or in binary systems) with low
mass-accretion rates are observed as much less luminous X-ray sources than the X-ray
transients. For example, the estimated X-ray emission from a single (non-binary)
black hole present within the interstellar medium is very small ($<
10^{34}~\mathrm{erg~s}^{-1}$; e.g., Ref.\citenum{2012MNRAS.427..589B}). It is
possible some of the extremely faint isolated stellar mass black holes have already
been found, but without hard band ($E > 10$~keV) coverage, it is impossible to
distinguish from other sources. Chandra performed a deep survey in the Galactic
Center region and detected 9,000 sources with $L_{\mathrm X} >
10^{31}~\mathrm{erg~s}^{-1}$\cite{2009ApJS..181..110M}. Although the majority of
these sources are thought to be white dwarf binaries\cite{2006A&A...452..169R}, some
should be StMBHs whose identifications are currently impossible. White dwarf
binaries have optically thin thermal emission with ~100 million K, whereas low
luminosity StMBHs must show power law spectra with a typical photon index of
$\sim$1.5 from the advection dominated accretion flows. To differentiate between
them, measuring broadband X-ray spectra that cover energies above 10~keV is
critical. The sensitivity of current instruments is limited to those with
$L_{\mathrm X} > 10^{33}~\mathrm{erg~s}^{-1}$, missing the bulk of the population
and making statistical studies impossible.


\subsection{How non-thermal energy is generated and how much non- thermal energy is
       contained in the universe}

According to the current particle acceleration paradigm, thermal particles with a
Maxwell-Boltzmann distribution attain high energies through diffusive shock
acceleration (or DSA, also known as first-order Fermi acceleration; see e.g.,
Ref.\citenum{1978MNRAS.182..147B}) which may occur at supernova shocks. This
paradigm has been successful in producing nonthermal, power-law energy spectra for
the distribution of accelerated particles, consistent with high-energy spectral
measurements in a variety of astrophysical systems, including SNRs. However, key
details about the DSA process have remained elusive for decades, particularly
whether and to what extent thermal particles are accelerated to nonthermal energies,
and how efficient the process is. 

In this regard, it is quite important to measure the total amount and the spectra of
sub-relativistic particles. Nonthermal bremsstrahlung in the hard X-ray energy range
is a promising emission channel to probe such particles. Recently, a NuSTAR
observation revealed hard X-ray emission in the SNR W49B, which was attributed to
nonthermal bremsstrahlung from electrons\cite{2018ApJ...866L..26T}. Such
observations are already challenging for the DSA process, as they pointed to either
the presence of a population of sub-relativistic particles or much more efficient
particle acceleration than predicted in the DSA paradigm. However, with its
backgrounds and angular resolution, NuSTAR was unable to provide precise
measurements of the amount of accelerated particles emitting in this band or their
spectrum. Both lower detector backgrounds and more efficient exclusion of cosmic
X-ray background are necessary to more precisely measure the spectrum of the
particles emitting in the hard X-ray band, particularly at the transition from the
thermal part of the spectrum to the nonthermal part that is inaccessible to
instruments less sensitive in the hard X-ray band. Detailed spatial information for
the hard X-ray emission is also important for comparison with maps of the ambient
gas. This will prove essential for translating the luminosity of the nonthermal
bremsstrahlung to the total amount of particles radiating in that band. 

In clusters of galaxies, particle acceleration may be studied in a manner similar to
the methods outlined for SNRs. From radio observations of synchrotron emission, we
know that relativistic particles with energies about 10$^{9}$~eV have to be
accelerated even in outer regions of clusters because their life is on the order of
10$^{9}$~yr. In order to separately constrain the magnetic fields and the energy
density of particles in intracluster environments, measurements of the energy
spectrum and image of inverse Compton emission from clusters of galaxies are
essential.  Inverse Compton emission arises when relativistic electrons (with
energies of 10$^{9}$~eV) scatter 2.7~K microwave background photons and raise the
photon energy to X-ray energies. The measurement needs to be done above
$\sim$10~keV, where the inverse Compton component is stronger than the thermal
emission of the intracluster gas. These data will tell us the energy density and
spectrum of electrons. The combination of the hard X-ray and radio data will solve
the degeneracy between magnetic field and electrons, giving us the magnetic field
strength without further assumption. 


There are a few clusters of galaxies with exceptionally high temperature of about
25~keV detected with Chandra and Suyaev-Zel'dovich measurements: RX
J1347.5$-$1145\cite{2008A&A...491..363O} ($z$=0.451) and
1E0657$-$56\cite{2002ApJ...567L..27M} ($z$=0.296). Existence of very hot clusters at
a fairly distant Universe can be a challenge to the cold dark matter scenario of
structure formation (e.g., Ref.\citenum{1998ApJ...496L...5T}), since the bottom-up
process can build up the mass to a certain limit within the cosmological time
scale. The high sensitivity observations in the high energy X-ray band will enable
us to determine ICM temperatures precisely for a wide range of clusters, and the
current scenario of the structure formation will be critically examined based on a
sample of the hottest clusters.

\subsection{How stars evolve and explode}

Type Ia supernova explosions are caused either by mass accretion of a white dwarf
from a stellar companion or merger of two white dwarfs forming a binary system. In
both scenarios, one of the important questions is whether supernovae occur with the
exploding white dwarf mass close to the Chandrasekhar limit ($\sim$ 1.4~\Msun) or
well below the limit. One of the most powerful diagnostic tools for disentangling
the two cases is nickel-to-iron (Ni/Fe) mass ratio\cite{2017ApJ...834..124Y,
2017Natur.551..478H} . In the case of the Chandrasekhar limit case, the core of the
white dwarf can be dense enough for the electron capture process to efficiently take
place, resulting in more abundant neutron-rich species including Ni. The Ni/Fe mass
ratio can be measured using K-shell transition lines of the elements, which appear
at $\sim$ 6--8 keV.

Recent theoretical studies suggest that asymmetry is a key ingredient to make
core-collapse supernovae explode\cite{2003ApJ...584..971B, 2014ApJ...786...83T,
2015ApJ...801L..24M, 2019MNRAS.482..351V}. Therefore, measurements of the asymmetry
can strongly constrain the models for core-collapse supernova explosions. We aim to
do so based on observations of line emissions from radioactive decay of $^{44}$Ti at
68~keV and 78~keV. Synthesized in the innermost part of supernova ejecta, both the
total amount synthesized and distribution of $^{44}$Ti serves as one of the most powerful
probes of asymmetry of core-collapse supernovae.

\section{Mission design overview}

\subsection{Mission requirement on angular resolution}
\label{sec:Requirement}

\begin{figure} [ht]
 \begin{minipage}{0.49\hsize}
  \centering \includegraphics[width=0.9\textwidth]{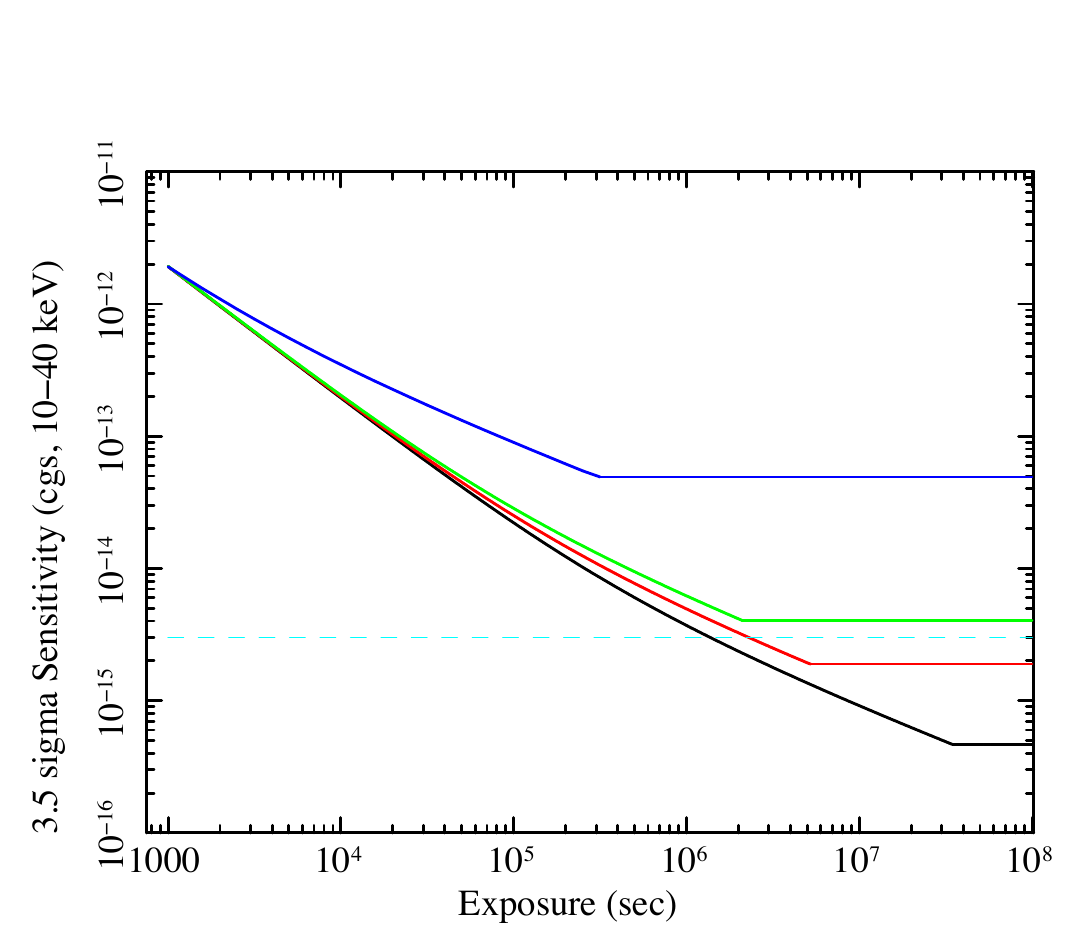}
  \caption[]{\label{fig:sensitivity} Point source sensitivity as a function of
  exposure time in deep surveys for CTAGNs. Black, red, green colors indicate those
  for $10'', 15''$, and $20''$ HPD cases, respectively, while blue color shows the
  Hitomi case. A cyan dashed line indicates the scientific requirement described in
  Sec.~\ref{sec:AGN}.}
 \end{minipage}
 \hspace{0.02\hsize}
  \begin{minipage}{0.49\hsize}
   \centering \includegraphics[width=0.9\textwidth]{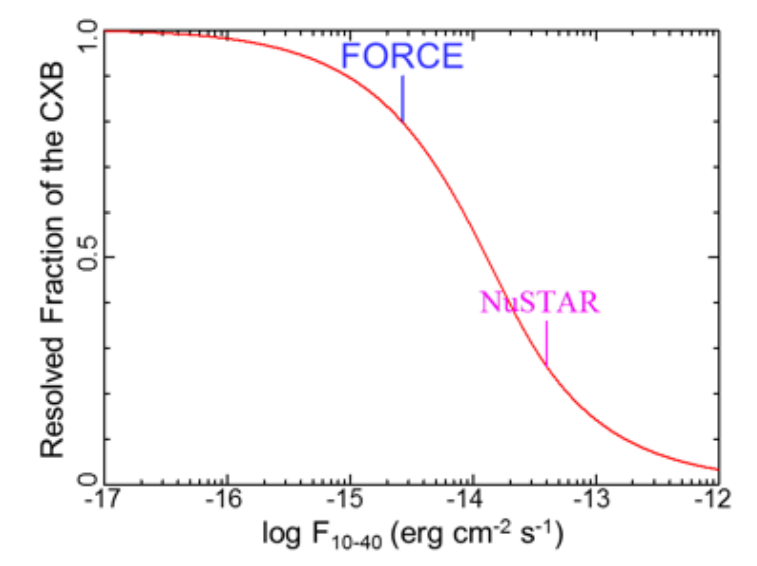}
   \caption[]{\label{fig:resolvedFraction} Resolved fraction of the CXB in the 10--40
   keV band as a function of flux limit predicted by a standard population synthesis
   model\cite{2014ApJ...786..104U}.}
 \end{minipage}
\end{figure}

The scientific objectives described in the previous section require unprecedented
high sensitivity in the hard X-ray band above 10~keV for both point and diffuse
sources. The most stringent requirement on sensitivity comes from the deep surveys
for CTAGNs. The deep surveys need highest sensitivity for the faintest point
sources, obscured and distant AGNs. As a bi-product, detection/exclusion of such
sources effectively reduces systematic uncertainly in the background estimation for
diffuse sources due to CXB fluctuation, resulting in higher sensitivity for diffuse
sources. Therefore, roughly speaking, all other science requirements on sensitivity
are met by the requirement from the deep surveys for CTAGNs.

As described in Sec.~\ref{sec:AGN}, a sensitivity limit of $3 \times
10^{-15}~\textrm{erg}~\textrm{cm}^{-2}~\textrm{s}^{-1}$ in the 10--40~keV band is
the threshold defined by the deep surveys for CTAGNs. In order to reach this limit,
the most important design parameter is angular
resolution. Fig.~\ref{fig:sensitivity} shows sensitivity curves (sensitivity reached
at a given exposure time) with several angular resolution cases. Here, the half
power diameter (HPD) of the point spread function (PSF) is used as a measure of
angular resolution. In any cases, in the short exposure regime less than 100~ks, the
sensitivity increases as the exposure time increases. However, at some point, the
sensitivity never gets better due to confusion of faint sources no mater how much
exposure time is invested. The sensitivity limit of $3 \times
10^{-15}~\textrm{erg}~\textrm{cm}^{-2}~\textrm{s}^{-1}$ requires that the angular
resolution must be better than 15$''$. Fig.~\ref{fig:resolvedFraction} shows the
resolved fraction of the CXB in the 10--40 keV band as a function of flux limit
predicted by a standard population synthesis model\cite{2014ApJ...786..104U}. The
FORCE's sensitivity limit makes it possible to resolve $\sim$ 80\% of the hard CXB
into discrete sources, meaning that we will be able to finally perform a thorough
census of all the accreting SMBHs in the universe. At the same time, we will be able
to understand the role of major mergers for SMBH growth, and hence the origin of the
coevolution.

\subsection{Mission and instrument design}

\begin{figure} [ht]
  \centering \includegraphics[width=0.6\textwidth]{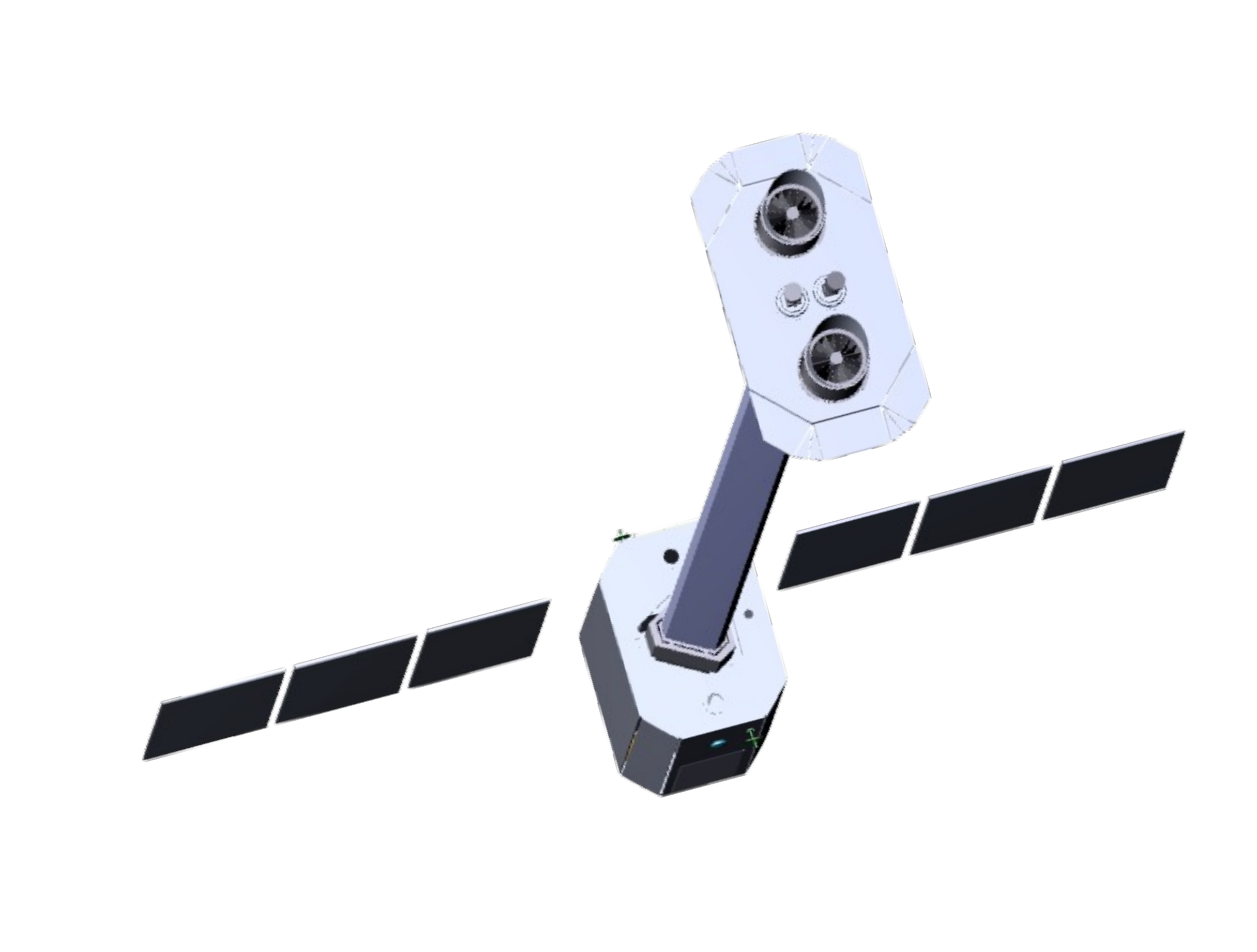}
  \caption[]{\label{fig:FORCE} Schematic view of the FORCE satellite.}
\end{figure}

\begin{table}[htbp]
  \centering
 \caption{ \label{tbl:param} Instrument parameters}
 \begin{tabular}{|l|l|}
  \hline
  Angular resolution (HPD) & $<$15$''$ \\
    \hline
  Multi-layer Coating & Pt/C \\
    \hline
  Field of view (50\% response) at 30~keV & $>$7$'$$\times$7$'$  \\
    \hline
  Effective Area at 30~keV & 230 cm$^{2}$ \\
  \hline
  Energy range & 1--79~keV \\
    \hline
  Energy resolution (FWHM) at 6~keV & $<$300~eV \\
    \hline
  Instrument background & comparable to those of Hitomi HXI\cite{2018JATIS...4b1409H} \\
    \hline
  Timing resolution (absolute) & several $\times$ 10~$\mu$s \\
  \hline
 \end{tabular}
\end{table}

Fig.~\ref{fig:FORCE} shows a schematic view of the FORCE satellite, and
Table~\ref{tbl:param} summarizes key instrument parameters. The overall mission
design is determined by scientific requirements, technical heritage of our previous
mission, \Hitomi, and restrictions from the launch vehicle. The weight of the
FORCE satellite is about 1 metric ton, and is planned to be launched into a circular
orbit with altitude of 500--600~km and inclination angle of 31~degrees or less by an
ISAS/JAXA solid-propellant Epsilon-S rocket. FORCE carries co-aligned, two identical pairs
of a supermirror with high angular resolution and a focal-plane detector with
broadband X-ray response. The supermirror and detector are separated by a focal
length of 12~m. The long focal length is essential to keep sufficient effective area
for hard X-ray, requiring an extendable optical bench (EOB) that can be stowed to
fit in the launch fairing and deployed on-orbit. Lateral displacements of the
detector in the alignment plane due to the thermal distortion of the EOB are
monitored by a laser metrology system\cite{Gallo14} in instantaneous fashion, and
the position reconstruction of detected X-rays are performed on ground on a
photon-by-photon basis. The precision of this system affects the net angular
resolution we finally obtain. In the case of \Hitomi, the movement
in the alignment plane was less than 400~$\mu$m with a focal length of
12~m\cite{Takahashi16}. We fully utilize the Hitomi heritages in our system. 


\subsubsection{X-ray supermirror based on light-weight, high-resolution silicon mirror}

The mirror substrates of our X-ray supermirrors are made based on the single-crystal
silicon mirror technology, which has been in development at NASA's Goddard Space
Flight Center\cite{ZhangSPIE2022}. X-ray mirror technologies are generally evaluated
from three aspects: angular resolution, mass per unit effective area, and cost per
unit effective area. Especially, budgetary-limited small missions need to find the
best compromise among these aspects, or rather focus on the latter two. The
single-crystal silicon mirror technology simultaneously meets the three-fold
requirement. Maturing processes toward a flight-ready X-ray mirror assembly are
extensively tested. 

\begin{figure} [ht]
 \begin{minipage}{0.49\hsize}
  \centering \includegraphics[width=0.9\textwidth]{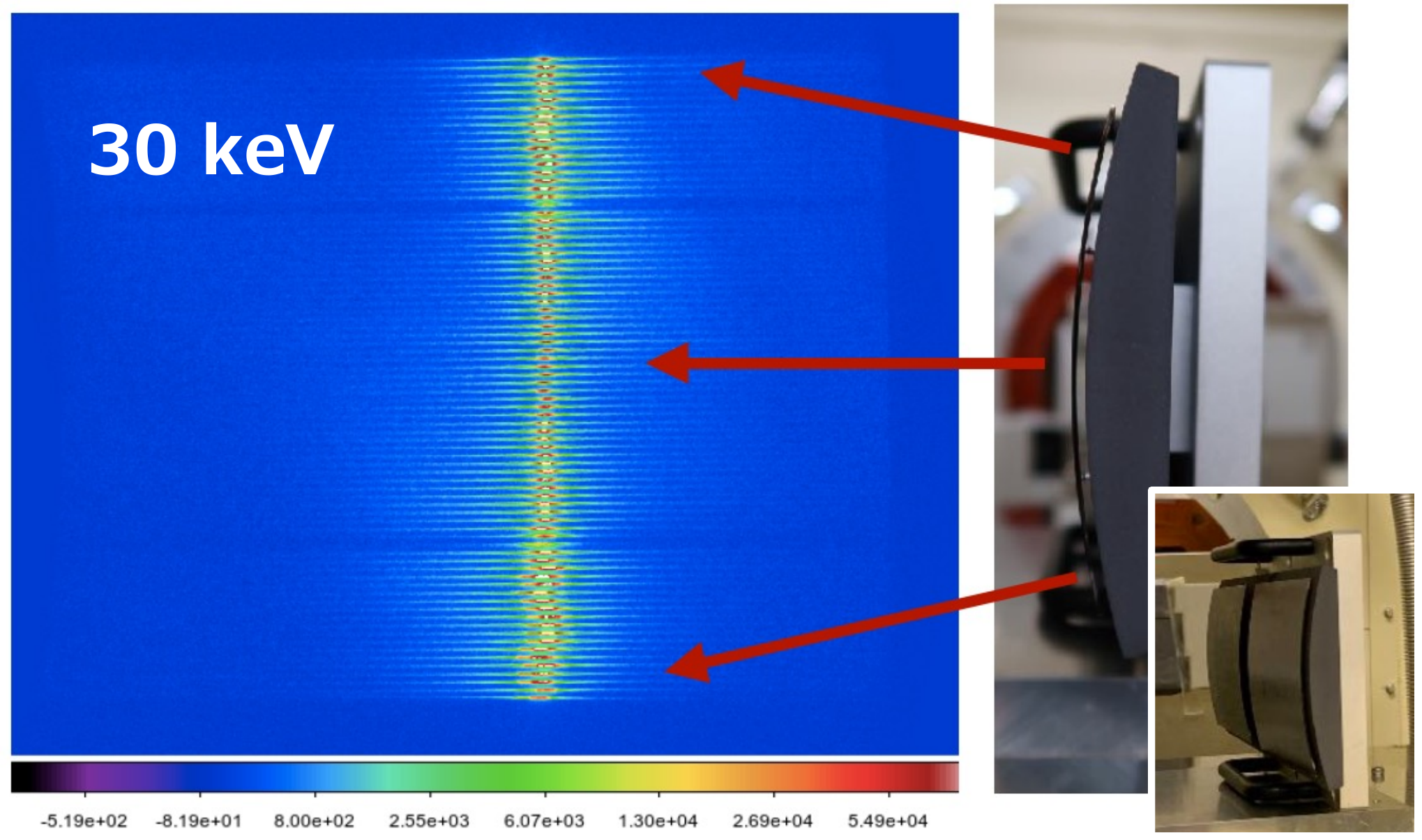}
  \caption[]{\label{fig:RasterScan} Reflected X-ray images obtained in a raster scan
  measurement with 30~keV X-ray beam. A silicon mirror module of a pair of the
  primary and secondary segments was tested, whose picture is shown in the inset.}
 \end{minipage}
 \hspace{0.02\hsize}
  \begin{minipage}{0.49\hsize}
   \centering \includegraphics[width=0.9\textwidth]{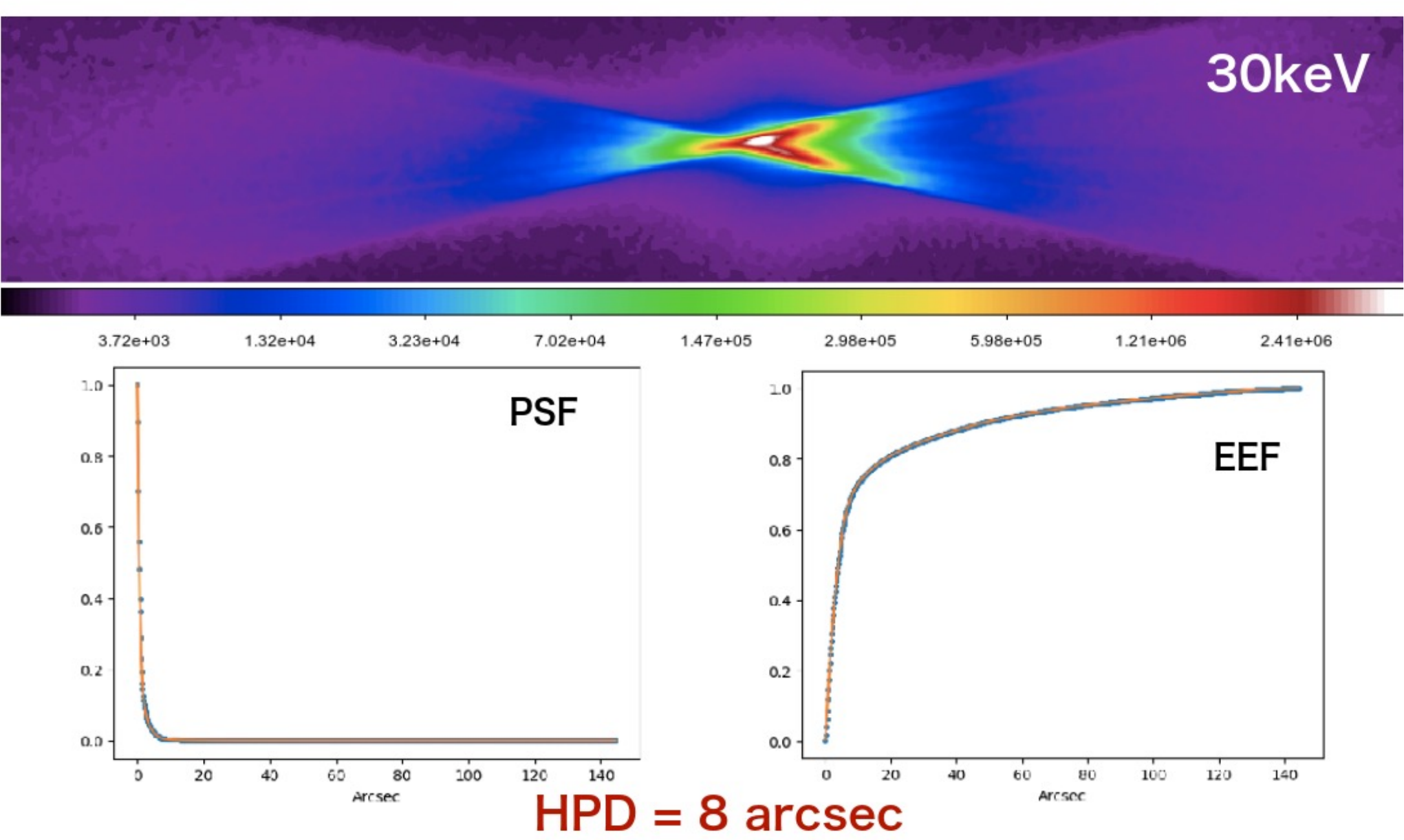}
   \caption[]{\label{fig:PSFandEEF} Summed image obtained from the raster scan
   measurement shown in Fig.~\ref{fig:RasterScan} (top). The PSF (bottom left) and
   EEF (bottom right) are derived from the summed image.}
 \end{minipage}
\end{figure}

We have been measuring X-ray performances of a multi-layer coated single-crystal
silicon mirror. FORCE will use the Wolter-I optical prescription, in which the
primary mirror is parabolic and the secondary mirror is hyperbolic in shape. A
silicon mirror module of a pair of the primary and secondary segments was made for
test purpose as shown in Fig.~\ref{fig:RasterScan} inset. The mirrors were coated
with Pt/C depth-graded multi-layers following a method used for the \Hitomi\ hard
X-ray mirror\cite{2018JATIS...4a1209T}. We carried out the characterization of the
module at the synchrotron facility SPring-8 BL20B2, where we can illuminate the
module with a monochromatic, hard X-ray parallel beam. Fig.~\ref{fig:RasterScan}
shows the result of a raster scan measurement with the 30~keV X-ray beam. The
reflected X-ray images slightly differ according to position, but are well
sorted. Fig.~\ref{fig:PSFandEEF} shows a summed image, and the PSF and encircled
energy function (EEF) derived from the summed image. The HPD of $\sim 8''$ was
obtained in this measurement. Although this result is from a single pair of the
primary and secondary segments, this value is well below the mission requirement.

\subsubsection{Wideband hybrid X-ray imager}


\begin{figure} [ht]
  \centering \includegraphics[width=0.25\textwidth]{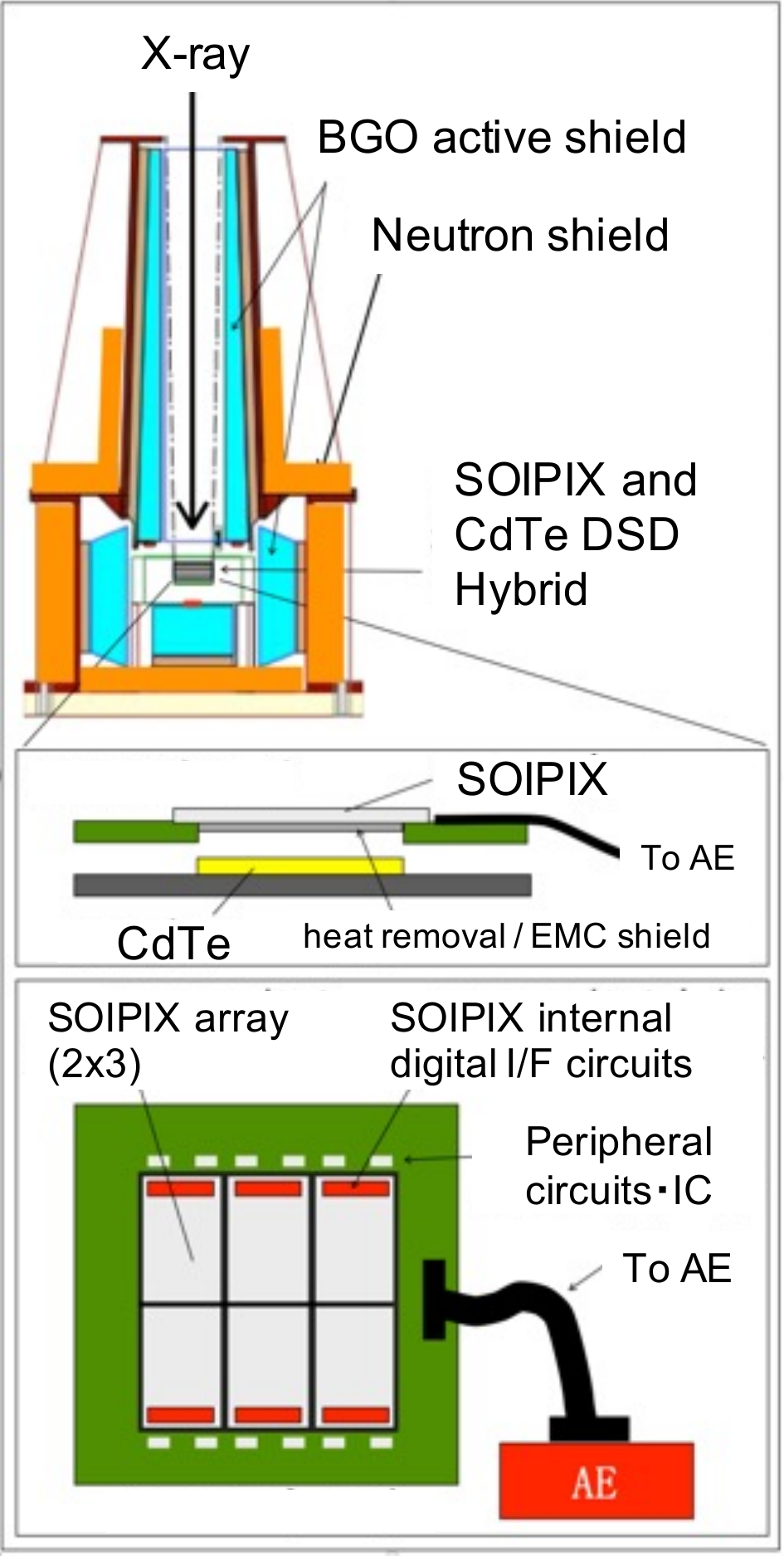}
  \caption[]{\label{fig:WHXI} Schematic drawing of WHXI}
\end{figure}

Fig.~\ref{fig:WHXI} shows a schematic drawing of the FORCE's focal plane detector,
Wideband Hybrid X-ray Imager (WHXI). The WHXI is a descendant of the Hard X-ray
Imager (HXI) onboard \Hitomi\cite{2018JATIS...4b1410N}, sharing a concept that Si
and CdTe hybrid detector surrounded by BGO active shield. The upper Si sensor is
responsible for incoming soft X-ray detection whereas the lower CdTe sensor receives
hard X-rays passing through the Si sensor, realizing broadband X-ray response. What
is new in the WHXI is to use a single layer of the SOI-CMOS pixel detector
(SOIPIX)\cite{TsuruSPIE2022} for the Si sensor in place of four layers of the
double-sided Si strip detector (DSSD) used in the HXI. Lower readout noise and
better energy resolution of SOIPIX in comparison with the DSSD can lower the energy
threshold down to 1~keV and allow us to perform X-ray line diagnostics in the Fe
band around 6~keV. SOIPIX is also equipped with a self-trigger function and fast
response of $<$ 10~$\mu$s so that the anti-coincidence technique can be applied as
in the case of the DSSD in the HXI. A single SOIPIX chip has a dimension of 22~mm
$\times$ 14~mm. In the WHXI, six SOIPIX chips are arranged in a 2$\times$3 grid,
resulting in an imaging area size of 44~mm $\times$ 44~mm with gaps between chips as
shown in Fig.~\ref{fig:WHXI}. With a focal length of 12~m, this number corresponds
to a $12.\!^{\prime}5 \times 12.\!^{\prime}5$ field of view. The CdTe DSD has a
dimension of 30~mm $\times$ 30~mm, covering a $8.\!^{\prime}5 \times 8.\!^{\prime}5$
field of view. Since the supermirror has an energy-dependent vignetting function and
provides a larger field of view in lower energy band, soft X-ray responsible SOIPIX
has a larger spatial coverage. We plan to operate the WHXI with a working
temperature of $-$25$\pm$1 $^{\circ}$C.

\begin{figure} [ht]
  \centering \includegraphics[width=0.9\textwidth]{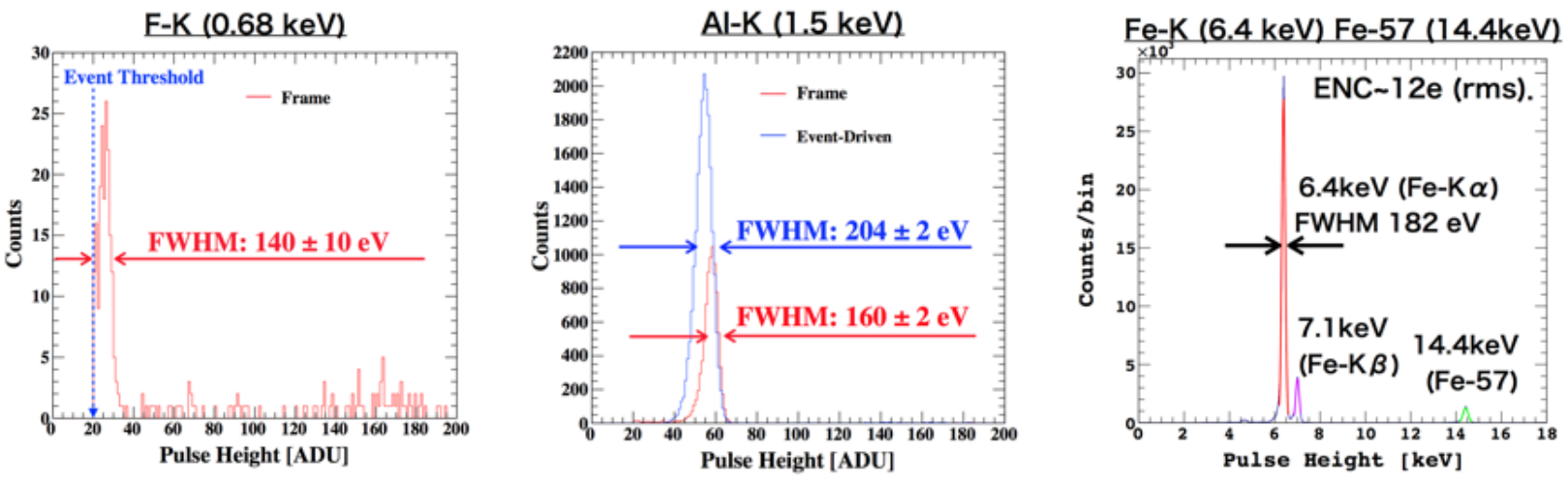}
  \caption[]{\label{fig:SOIPIX-spectra} X-ray spectra taken with SOIPIX at 0.68~keV
 (left), 1.5~keV (middle), and 6.4~keV (right). Note that the spectrum at 0.68~keV
 is taken with the frame mode, not with the event-driven mode\cite{KODAMA2021164745}.}
\end{figure}

Fig.~\ref{fig:SOIPIX-spectra} shows the current X-ray spectroscopic performance of
SOIPIX. Please note that these spectra were taken with a small size
chip. Nonetheless, the energy resolution at 6.4 keV is close to that of X-ray
CCD. Al-K X-rays at 1.5~keV were successfully detected with the event-driven
readout, but F-K X-rays at 0.68~keV were not\cite{KODAMA2021164745}. Further
studies in the reduction of readout noise are ongoing. 

\section{Roles in multi-messenger astronomy era in the 2030s}

\begin{figure} [ht]
  \centering \includegraphics[width=0.6\textwidth]{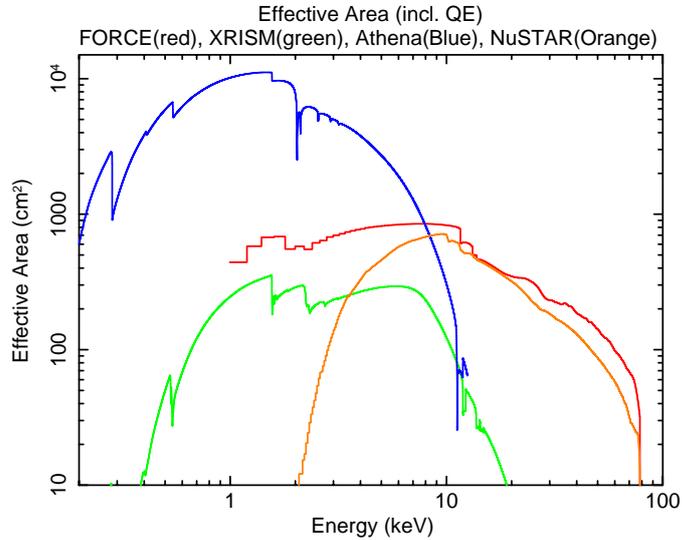}
  \caption[]{\label{fig:FORCE-Athena} Comparison of effective area among FORCE (red),
 NuSTAR (orange), XRISM (green), and Athena (blue).}
\end{figure}

Fig.~\ref{fig:FORCE-Athena} shows an effective area of FORCE as well as those of
NuSTAR, XRISM, and Athena. In comparison with NuSTAR, FORCE is characterized by an
enhancement of the effective area in soft X-ray band below 10~keV, providing
simultaneous broad band coverage without coordinated observations with other
observatories. XRISM, Japan's seventh X-ray astronomy mission to be launched in the
Japanese fiscal year 2022, is expected to revolutionize our understanding of the
X-ray Universe with its unprecedented energy resolution. Athena will follow and
expand the high resolution spectroscopy in 2030s. The Athena design is optimized to
have a large effective area in the soft X-ray band around 1~keV, which is in clear
contrast with that of FORCE, and therefore FORCE and Athena are quite complementary
to each other in 2030s.

Hard X-ray studies are also complementary to current and future gravitational wave
studies. Gravitational waves provide a useful way to find IMBHs that grow via
mergers. LIGO, Virgo, and KAGRA will detect merging objects and are sensitive to the
lower end of the IMBH range (a few hundred solar masses). LISA will sample the upper
range ($> 10^{4}$ \Msun) of IMBHs. FORCE fills in this mass gap with sensitivity
over the entire range and to accreting sources (growing in a different way than
mergers). Accreting IMBHs, observed as ULX or HLX sources, may only be discovered in
the X-ray band, where most of their emission is radiated. LISA will also observe a
significant population of Milky Way compact object binaries with wide orbits that
will include StMBH. However, it will be heavily biased towards mostly white dwarf
systems (e.g., Ref.~\citenum{Breivik_2018}). This LISA population of wide-orbit
StMBHs will complement the FORCE population of isolated/quiescent BHs in the
Galactic center to complete the full picture of StMBHs. 

FORCE will also be a strong complement to TeV and neutrino observatories. Observing
gamma rays in TeV energies, the Cherenkov Telescope Array (CTA) can also probe
ultra-relativistic particles but with neutral pion decay emission, a different
radiation channel from what we aim to detect with FORCE. In addition, future
neutrino detectors would be able to detect high energy neutrino emission from SNRs,
which are the decay products of charged pions. Hard X-ray data, therefore,
constitute an important piece in the new field of multi-messenger
astronomy. Combining hard X-ray, TeV gamma-ray, and neutrino data, we can solidly
constrain the maximum acceleration energy in SNRs and can give a conclusive answer
to the question about the origin of Galactic cosmic rays.

\section{SUMMARY}
\label{sec:summary}

We here present the Focusing on Relativistic universe and Cosmic Evolution (FORCE)
mission, the product of a JAXA/NASA collaboration. The FORCE mission will achieve 10
times higher sensitivity in the hard X-ray band in comparison to any previous hard
X-ray mission. FORCE aims to be launched in the early 2030s, as a perfect hard X-ray
complement to Athena. FORCE provides broadband (1--79 keV) X-ray imaging
spectroscopy with high angular resolution ($< 15''$). FORCE will be the most
powerful X-ray probe for discovering obscured/hidden black holes and studying high
energy particle acceleration in our Universe.

\acknowledgments 

The measurements of the X-ray mirror for FORCE were done at SPring-8 2018B1172,
2019B1274, and 2020A1285. This work is partially supported by JSPS KAKENHI Grant
Numbers JP21H01095, JP20H00157, JP22H04572, JP21H04493, JP21H05461, JP21K18151,
JP20H00175, JP19K14742.


\bibliography{mybibfile} 

\begin{thebibliography}{10}

\bibitem{2016SPIE.9905E..1OM}
{Mori}, K. et~al., ``{A broadband x-ray imaging spectroscopy with high-angular
  resolution: the FORCE mission},'' in [{\em Space Telescopes and
  Instrumentation 2016: Ultraviolet to Gamma Ray}{\nolinebreak\hspace{0.1em}]},
   {den Herder}, J.-W.~A., {Takahashi}, T., and {Bautz}, M., eds., {\em Society
  of Photo-Optical Instrumentation Engineers (SPIE) Conference Series} {\bf
  9905},  99051O (July 2016).

\bibitem{2018SPIE10699E..2DN}
{Nakazawa}, K. et~al., ``{The FORCE mission: science aim and instrument
  parameter for broadband x-ray imaging spectroscopy with good angular
  resolution},'' in [{\em Space Telescopes and Instrumentation 2018:
  Ultraviolet to Gamma Ray}{\nolinebreak\hspace{0.1em}]},  {den Herder},
  J.-W.~A., {Nikzad}, S., and {Nakazawa}, K., eds., {\em Society of
  Photo-Optical Instrumentation Engineers (SPIE) Conference Series} {\bf
  10699},  106992D (July 2018).

\bibitem{2009ApJ...698..198G}
{G{\"u}ltekin}, K. et~al., ``{The M-{\ensuremath{\sigma}} and M-L Relations in
  Galactic Bulges, and Determinations of Their Intrinsic Scatter},'' {\em
  \apj}~{\bf 698},  198--221 (June 2009).

\bibitem{2017ApJS..233...17R}
{Ricci}, C. et~al., ``{BAT AGN Spectroscopic Survey. V. X-Ray Properties of the
  Swift/BAT 70-month AGN Catalog},'' {\em \apjs}~{\bf 233},  17 (Dec. 2017).

\bibitem{2016ApJ...831..185H}
{Harrison}, F.~A. et~al., ``{The NuSTAR Extragalactic Surveys: The Number
  Counts of Active Galactic Nuclei and the Resolved Fraction of the Cosmic
  X-Ray Background},'' {\em \apj}~{\bf 831},  185 (Nov. 2016).

\bibitem{2015ApJ...808..185C}
{Civano}, F. et~al., ``{The Nustar Extragalactic Surveys: Overview and Catalog
  from the COSMOS Field},'' {\em \apj}~{\bf 808},  185 (Aug. 2015).

\bibitem{2017ApJ...846...20L}
{Lansbury}, G.~B. et~al., ``{The NuSTAR Serendipitous Survey: Hunting for the
  Most Extreme Obscured AGN at $>$10 keV},'' {\em \apj}~{\bf 846},  20 (Sept.
  2017).

\bibitem{2000ApJ...535..632M}
{Makishima}, K. et~al., ``{The Nature of Ultraluminous Compact X-Ray Sources in
  Nearby Spiral Galaxies},'' {\em \apj}~{\bf 535},  632--643 (June 2000).

\bibitem{2014Natur.514..202B}
{Bachetti}, M. et~al., ``{An ultraluminous X-ray source powered by an accreting
  neutron star},'' {\em \nat}~{\bf 514},  202--204 (Oct. 2014).

\bibitem{2016ApJ...831L..14F}
{F{\"u}rst}, F. et~al., ``{Discovery of Coherent Pulsations from the
  Ultraluminous X-Ray Source NGC 7793 P13},'' {\em \apjl}~{\bf 831},  L14 (Nov.
  2016).

\bibitem{2011MNRAS.416.1844W}
{Walton}, D.~J. et~al., ``{2XMM ultraluminous X-ray source candidates in nearby
  galaxies},'' {\em \mnras}~{\bf 416},  1844--1861 (Sept. 2011).

\bibitem{1996ApJ...473..963B}
{Barret}, D. et~al., ``{Luminosity Differences between Black Holes and Neutron
  Stars},'' {\em \apj}~{\bf 473},  963 (Dec. 1996).

\bibitem{1998ApJ...496..155S}
{Samland}, M., ``{Modeling the Evolution of Disk Galaxies. II. Yields of
  Massive Stars},'' {\em \apj}~{\bf 496},  155--171 (Mar. 1998).

\bibitem{2016PASJ...68S...1N}
{Negoro}, H. et~al., ``{The MAXI/GSC Nova-Alert System and results of its first
  68 months},'' {\em \pasj}~{\bf 68},  S1 (June 2016).

\bibitem{2012MNRAS.427..589B}
{Barkov}, M.~V., {Khangulyan}, D.~V., and {Popov}, S.~B., ``{Jets and gamma-ray
  emission from isolated accreting black holes},'' {\em \mnras}~{\bf 427},
  589--594 (Nov. 2012).

\bibitem{2009ApJS..181..110M}
{Muno}, M.~P. et~al., ``{A Catalog of X-Ray Point Sources from Two Megaseconds
  of Chandra Observations of the Galactic Center},'' {\em \apjs}~{\bf 181},
  110--128 (Mar. 2009).

\bibitem{2006A&A...452..169R}
{Revnivtsev}, M. et~al., ``{Origin of the Galactic ridge X-ray emission},''
  {\em \aap}~{\bf 452},  169--178 (June 2006).

\bibitem{1978MNRAS.182..147B}
{Bell}, A.~R., ``{The acceleration of cosmic rays in shock fronts - I.},'' {\em
  \mnras}~{\bf 182},  147--156 (Jan. 1978).

\bibitem{2018ApJ...866L..26T}
{Tanaka}, T. et~al., ``{NuSTAR Detection of Nonthermal Bremsstrahlung from the
  Supernova Remnant W49B},'' {\em \apjl}~{\bf 866},  L26 (Oct. 2018).

\bibitem{2008A&A...491..363O}
{Ota}, N. et~al., ``{Suzaku broad-band spectroscopy of RX J1347.5-1145:
  constraints on the extremely hot gas and non-thermal emission},'' {\em
  \aap}~{\bf 491},  363--377 (Nov. 2008).

\bibitem{2002ApJ...567L..27M}
{Markevitch}, M. et~al., ``{A Textbook Example of a Bow Shock in the Merging
  Galaxy Cluster 1E 0657-56},'' {\em \apjl}~{\bf 567},  L27--L31 (Mar. 2002).

\bibitem{1998ApJ...496L...5T}
{Tucker}, W. et~al., ``{1E 0657-56: A Contender for the Hottest Known Cluster
  of Galaxies},'' {\em \apjl}~{\bf 496},  L5--L8 (Mar. 1998).

\bibitem{2017ApJ...834..124Y}
{Yamaguchi}, H. et~al., ``{The Origin of the Iron-rich Knot in
  Tycho{\textquoteright}s Supernova Remnant},'' {\em \apj}~{\bf 834},  124
  (Jan. 2017).

\bibitem{2017Natur.551..478H}
{Hitomi Collaboration}, ``{Solar abundance ratios of the iron-peak elements in
  the Perseus cluster},'' {\em \nat}~{\bf 551},  478--480 (Nov. 2017).

\bibitem{2003ApJ...584..971B}
{Blondin}, J.~M., {Mezzacappa}, A., and {DeMarino}, C., ``{Stability of
  Standing Accretion Shocks, with an Eye toward Core-Collapse Supernovae},''
  {\em \apj}~{\bf 584},  971--980 (Feb. 2003).

\bibitem{2014ApJ...786...83T}
{Takiwaki}, T., {Kotake}, K., and {Suwa}, Y., ``{A Comparison of Two- and
  Three-dimensional Neutrino-hydrodynamics Simulations of Core-collapse
  Supernovae},'' {\em \apj}~{\bf 786},  83 (May 2014).

\bibitem{2015ApJ...801L..24M}
{Melson}, T., {Janka}, H.-T., and {Marek}, A., ``{Neutrino-driven Supernova of
  a Low-mass Iron-core Progenitor Boosted by Three-dimensional Turbulent
  Convection},'' {\em \apjl}~{\bf 801},  L24 (Mar. 2015).

\bibitem{2019MNRAS.482..351V}
{Vartanyan}, D. et~al., ``{A successful 3D core-collapse supernova explosion
  model},'' {\em \mnras}~{\bf 482},  351--369 (Jan. 2019).

\bibitem{2014ApJ...786..104U}
{Ueda}, Y. et~al., ``{Toward the Standard Population Synthesis Model of the
  X-Ray Background: Evolution of X-Ray Luminosity and Absorption Functions of
  Active Galactic Nuclei Including Compton-thick Populations},'' {\em
  \apj}~{\bf 786},  104 (May 2014).

\bibitem{2018JATIS...4b1409H}
{Hagino}, K. et~al., ``{In-orbit performance and calibration of the Hard X-ray
  Imager onboard Hitomi (ASTRO-H)},'' {\em Journal of Astronomical Telescopes,
  Instruments, and Systems}~{\bf 4},  021409 (Apr. 2018).

\bibitem{Gallo14}
{Gallo}, L. et~al., ``{The Canadian Astro-H Metrology System},'' in [{\em Space
  Telescopes and Instrumentation 2014: Ultraviolet to Gamma
  Ray}{\nolinebreak\hspace{0.1em}]},   {\bf 9144},  914456 (2014).

\bibitem{Takahashi16}
{Takahashi}, T. et~al., ``{The ASTRO-H x-ray astronomy satellite},'' in [{\em
  Astronomical Telescopes and Instruments 2016
  Proceeding}{\nolinebreak\hspace{0.1em}]},  {\em Proc. SPIE} (2016).

\bibitem{ZhangSPIE2022}
{Zhang}, W.~W. et~al., ``{Single Crystal Silicon X-ray Optics for Astronomy:
  high resolution, light weight, and low cost},'' {\em \it\procspie},  in press
  (2022).

\bibitem{2018JATIS...4a1209T}
{Tamura}, K. et~al., ``{Supermirror design for Hard X-Ray Telescopes on-board
  Hitomi (ASTRO-H)},'' {\em Journal of Astronomical Telescopes, Instruments,
  and Systems}~{\bf 4},  011209 (Jan. 2018).

\bibitem{2018JATIS...4b1410N}
{Nakazawa}, K. et~al., ``{Hard x-ray imager onboard Hitomi (ASTRO-H)},'' {\em
  Journal of Astronomical Telescopes, Instruments, and Systems}~{\bf 4},
  021410 (Apr. 2018).

\bibitem{TsuruSPIE2022}
{Tsuru}, T.~G. et~al., ``{Recent Progress in Development of Trigger-Output
  Event-Driven X-ray astronomy SOI pixel sensors},'' {\em \it\procspie},  in
  press (2022).

\bibitem{KODAMA2021164745}
{Kodama}, R. et~al., ``Low-energy x-ray performance of soi pixel sensors for
  astronomy, “xrpix”,'' {\em Nuclear Instruments and Methods in Physics
  Research Section A: Accelerators, Spectrometers, Detectors and Associated
  Equipment}~{\bf 986},  164745 (2021).

\bibitem{Breivik_2018}
{Breivik}, K. et~al., ``Characterizing accreting double white dwarf binaries
  with the laser interferometer space antenna and gaia,'' {\em The
  Astrophysical Journal}~{\bf 854},  L1 (feb 2018).

\end{thebibliography}
\bibliographystyle{spiebib} 

\end{document}